\documentclass[aps,prl,floats,showpacs,twocolumn]{revtex4}
\usepackage{graphicx}
\begin{document}
\title{Localized excitations at the Mott Insulator-Superfluid interfaces for confined Bose-Einstein condensates}
\author{Eros Mariani and Ady Stern
  \vspace{1mm}}
\affiliation{Department of Condensed Matter Physics \\
The Weizmann Institute of Science, 76100 Rehovot (Israel)
\vspace{3mm}}
\date{March 27, 2005}
\begin{abstract} 
In this paper we derive the dispersion relation of the surface waves at the interfaces between Mott
insulating and Superfluid phases for confined Bose-Einstein Condensates. We then calculate their contribution to the
heat capacity of the system and show how its low temperature scaling allows an experimental test of the existence and
properties of Mott Insulator-Superfluid domains.
\end{abstract}
\pacs{32.80.Pj, 03.75.Lm, 03.75.Hh, 03.75.Kk}
\maketitle
%%%%%%%%%%%%%%%%%%%%%%%%%%%%%%%%%%%%%%%%%%%%%%%%%%%%%%%%%%%%%%%%%%%%

Following the experimental observation of Bose-Einstein Condensation (BEC) \cite{Ketterle} a lot of attention has been
given
 to the investigation of the possible quantum phase transitions such a system can undergo. 
Within the framework of the Bose-Hubbard model, Fisher et al. \cite{Fisher89} proposed the zero temperature phase
diagram of an \emph{infinite uniform system on a lattice} as a function of the typical hopping kinetic energy $J$, the
atomic on-site repulsive interaction $U$ and their chemical potential $\mu$. In the $(J/U, \mu/U)$ plane they found
lobe-like phase boundaries between Mott-Insulating (MI) and Superfluid (SF) phases for small and large values of $J/U$,
respectively.\\
Recent experiments on BEC in a periodic optical lattice showed the expected MI-SF transition \cite{Orzel01,Greiner02}
driven by variations of the system parameters, while theoretical investigations focused on the restoration of macroscopic
phase coherence after abrupt parametric changes \cite{Altman,Polkovnikov}.\\
Real experiments are however always performed on \emph{finite} systems subject to an external confinement, yielding an
effective
space-dependent chemical potential. The consequent modulation in the local boson density can produce spatial-domains
characterized by different equilibrium phases (MI or SF). Indeed, it has been recently proposed that the observed MI-SF
crossover is to be interpreted as a relative widening/shrinking of the different phase domains rather than as a pure
infinite long-range transition \cite{Wessel04}.\\
Recent numerical calculations investigated the equilibrium state of the confined BEC and confirmed the existence of the
above-mentioned MI-SF domains \cite{Jaksch98,Kashurnikov02,Batrouni02}. 
However, their experimental signatures are still an open issue. 
Moreover, it is clear that the interfaces between the different phases can play an important role in the thermodynamic
and time-evolution properties of the system under external perturbations.\\
In this paper we consider the interfaces between MI and SF phases and evaluate the dispersion of their surface wave
excitations. We show that the low temperature heat capacity of the system is crucially affected by the
existence and properties of MI-SF domains, allowing for an experimental way to detect them.

A BEC at zero temperature in a periodic optical lattice can be described by the Bose-Hubbard Hamiltonian
\begin{eqnarray}
\label{H} H&=&-J\sum_{\langle ij\rangle}^{}\left(a_{i}^{\dagger}a_{j}^{}+h.c.\right)+\sum_{
i}^{}\left(-\mu+V_{i}^{}\right)a_{i}^{\dagger}a_{i}^{}+\nonumber \\
&+&\frac{U}{2}\sum_{i}^{}a_{i}^{\dagger}a_{i}^{}\left(a_{i}^{\dagger}a_{i}^{}-1\right)
\end{eqnarray}
with $a_{i}^{}$ the annihilation operator for a boson at site $i$ and $\langle ij\rangle$ indicates nearest-neighbor
hopping. 
The single-particle kinetic term $J$ can be tuned by the intensity of the light generating the optical lattice, the
on-site repulsion $U$ by the use of Feshbach resonances while the chemical potential $\mu$ depends on the particle
number and $V_{i}^{}$ describes an external confining potential.\\
For the uniform case $V_{i}^{}=0$, the system described by the Hamiltonian (\ref{H}) undergoes quantum phase transitions
between MI phases and SF ones, driven by the \emph{two} parameters $J/U$ and $\mu/U$. In the MI-phases the number of
particles per-site is fixed to a positive \emph{integer}, producing a constant density, in the continuum limit.\\ 
Under the influence of a confinement, the effective chemical potential $\mu-V_{i}^{}$ is position dependent and the
sample can show MI phases and SF ones in different domains. 
We will consider an effective two-dimensional system in the $(x,z)$ plane with a one-dimensional confinement along the
$z$-direction and translational invariance along the $x$-axis. A realistic cylindrically symmetric confinement reduces
to our model by remapping the radial direction into the $z$-axis and considering the $x$ direction as closed per
periodicity. In principle many successive alternated MI and SF phases can be formed, according to the particle density,
but we will mainly concentrate on one SF domain "guarded" by two neighboring MI ones. The confinement is such to produce
a MI-phase with fixed integer site occupation $n_{\mathrm{M}}^{}$ for $z<0$ and the next MI with occupation
$n_{\mathrm{M}}^{}+1$ for $z>L$, with a SF-phase with variable density for $0\le z\le L$. If we use, for simplicity, a
square lattice with typical lattice size $l$, in the continuum the two MI have densities
$\rho_{\mathrm{M}}^{}=n_{\mathrm{M}}^{}/l_{}^{2}$ and $\rho_{\mathrm{M}}^{}+\Delta \rho
=(n_{\mathrm{M}}^{}+1)/l_{}^{2}$. The equilibrium SF density $\bar{\rho}(z)$ varies continuously between these values, 
\begin{equation}
\label{rho}
\bar{\rho}(z)=\rho_{\mathrm{M}^{}}+ f(z)\,\Delta \rho
\end{equation}
where $f(z)$ is determined in principle by solving the Gross-Pitayevski equation in presence of the external confinement
and is such that $f(0)=0$ and $f(L)=1$.

We imagine to impose an infinitesimal displacement $\zeta (x,t)\propto \exp [i(kx-\omega t)]$, $\zeta \ll L$, along the
$z$-direction to the interface at $z=0$ while that at $z=L$ does not move. Our task will be to
determine the change in the "potential energy" associated to this perturbation via the surface tension of the interface, as well as
the "kinetic energy" due to the induced SF velocity. Up to quadratic terms in $\zeta$ and $\dot{\zeta}$, for any $k$, the
former contribution will have a generic
form $1/2\, M\omega^{2}_{}(k)\zeta^{2}_{}$ and the latter $1/2\, M\dot{\zeta}^{2}_{}$ with $M$ a generalized mass,
whence the surface waves dispersion $\omega (k)$.\\
This problem is similar to the determination of the dispersion for crystallization waves at a quantum interface between
solid and liquid He$^{4}_{}$ \cite{Andreev78}. However, the continuity of the density profile and the presence of the cutoff
length $L$ bring important new features into our case.

Let us first calculate the induced change in the potential energy of the interface
\begin{equation}
\label{U} U=\int \mathrm{d}l\, \alpha (\mathbf{n})\quad ,
\end{equation}
where d$l$ is the unit length element along the surface and $\alpha (\mathbf{n})$ is the surface energy per unit length
as a function of the normal $\mathbf{n}$ to the interface. In the $k=0$ case, the rigid shift of the interface away from
its equilibrium location will correspond to an energy cost $1/2\,
\Delta \left|\zeta \right|_{}^{2}$, whose harmonic confinement frequency $\Delta$ will be analyzed later.\\
For finite-$k$ and small $\zeta$ we can expand $n_{z}^{}\cong 1$ and $n_{x}^{}\cong \partial_{x}^{}\zeta$, so that 
$\mathrm{d}l\cong \mathrm{d}x\,\left[1+1/2\left(
\partial_{x}^{}\zeta\right)^{2}_{}\right]$. 
The surface tension term undergoes the expansion 
\begin{equation}
\label{alphan}
\alpha (\mathbf{n})\cong \alpha (\hat{z})+\frac{
\partial\alpha}{
\partial n_{x}^{}}\cdot
\partial_{x}^{}\zeta+\frac{1}{2}\,\frac{
\partial_{}^{2}\alpha}{
\partial n_{x}^{2}}\cdot
\left(
\partial_{x}^{}\zeta\right)^{2}_{}
\end{equation}
so that the \emph{variation} per unit-length in the potential energy of the interface due to the modulation is
\begin{equation}
\label{DeltaU}
\Delta U=\frac{1}{2}\left[\Delta + \left(\alpha (\hat{z})+\frac{\partial_{}^{2}\alpha}{
\partial n_{x}^{2}}\right)k^{2}_{}\right]\,\left|\zeta\right|^{2}_{}\quad ,
\end{equation}
where we used the relation $
\partial_{x}^{}\zeta =ik\zeta$.

Now let us concentrate on the kinetic energy associated to the superfluid velocity $\mathbf{v}$ induced by the surface
displacement. In order to determine the velocity profile we consider the continuity equation
\begin{equation}
\label{continEq}
\partial_{t}^{}\rho_{\mathrm{loc}}^{}(x,z,t)+\mathbf{\nabla}\cdot\left[\,\rho_{\mathrm{loc}}^{}(x,z,t)\mathbf{v}(x,z,t)\,
\right]=0
\end{equation}
with $\rho_{\mathrm{loc}}^{}(x,z,t)$ the local density. In the $k=0$ case, the density as well as the velocity do not
depend on $x$ and the only solution for $v$ (the $z$-component of $\mathbf{v}$) yielding a vanishing equilibrium current
at $z\ge L$ is
\begin{equation}
v(z,t)=\frac{1}{\rho_{\mathrm{loc}}^{}(z,t)}\int_{z}^{\infty}\mathrm{d}y\,
\partial_{t}^{}\rho_{\mathrm{loc}}^{}(y,t)\quad .
\end{equation}
The consequent uniform velocity in the MI phase at $z<0$ is needed to compensate for the loss (or addition) of particles
implied by the variation in the density profile. The related current running in the insulating domain 
is therefore associated to a very slow relaxation process.\\
In the finite-$k$ regime the two components of the velocity $\mathbf{v}=(v_{x}^{},v_{z}^{})$, show up. The SF constraint
$\nabla \times \mathbf{v}(x,z,t)=0$ is accounted for by defining $\mathbf{v}$ as a potential gradient, 
$\mathbf{v}(x,z,t)=\nabla
\psi(x,z,t)$.\\
In the limit of small oscillations, we write the density in terms of its equilibrium profile and a small induced
fluctuation, $\rho_{\mathrm{loc}}^{}(x,z,t)=\bar{\rho}(z)+\rho (x,z,t)$. 
Thus, $\rho$ and the related velocity $\mathbf{v}$ are the small quantities around which to expand the continuity
equation (\ref{continEq}) to obtain
\begin{equation}
\label{contToy3}
\partial_{z}^{2}\psi -k^{2}_{}\psi +
\partial_{z}^{}\left[\log \bar{\rho}\right]
\partial_{z}^{}\psi+\frac{\dot{\rho}}{\bar{\rho}}=0\quad .
\end{equation} 
The \emph{generic} features of the surface waves we consider can be grasped by a model where we assume a linear
dependence of the equilibrium density on the $z$ coordinate, i.e. in (\ref{rho}) we substitute $f(z)=z/L$. Such a
description is relevant for either
a linear external confinement or for the regions where a generic confinement can be effectively linearized. In any case,
corrections to this model due to the curvature of the realistic confinement should not alter the main
\emph{qualitative} effects we now investigate.\\
Eq. (\ref{contToy3}) can be solved in the limit $\rho_{\mathrm{M}}/\Delta \rho\gg 1$, (i.e. large site-occupation
$n_{\mathrm{M}}^{}\gg 1$). Then, 
$\partial_{z}^{}\left[\log\bar{\rho}\right]\cong 2q$,  
having defined $q=\Delta \rho/\left(2L\rho_{\mathrm{M}}\right)$, with the dimensions of a wave-number, and $qL\ll 1$. 
In the small occupation limit (always at least $n_{\mathrm{M}}^{}\ge 1$) this approximation is invalidated only close
to $z=L$
and no major 
\emph{qualitative} features will change.\\
In order to specify the inhomogeneous term $\dot{\rho}/\bar{\rho}$, we 
consider the dynamics via the linearized Newton's law
\begin{equation}
\label{Newton} 
m\bar{\rho}(z)\,
\partial_{t}^{}\mathbf{v}(x,z,t)+\frac{1}{\kappa}\mathbf{\nabla}\rho(x,z,t)=0
\end{equation}
with $m$ the mass of the single bosonic atom and $\kappa =
\partial \rho /
\partial \mu$ the compressibility of the superfluid, assumed constant for simplicity. Eq. (\ref{Newton}), together with
the linearized continuity equation, 
lead to the familiar wave equation 
\begin{equation}
\label{Wave}
\partial^{2}_{t}\rho (x,z,t)-c^{2}_{}\nabla ^{2}_{}\rho (x,z,t)=0
\end{equation}
yielding bulk sound waves with group velocity $c=(m\kappa )^{-1/2}_{}=
\left[2\pi\rho^{}_{\mathrm{M}} l^{2}_{}U/m\right]^{1/2}_{}$
\cite{Pitayevski}. 
The solution to the wave equation is of the form $\rho \propto \exp [i(kx-\omega t)\pm pz]$ with
$p^{2}_{}=k^{2}_{}(1-\omega ^{2}_{}/c^{2}_{}k^{2}_{})$, while 
the interface modulation corresponds to a boundary condition $\rho (x,z=0,t)=-2q\rho_{\mathrm{M}}^{}\zeta (x,t)$ and we
still preserve 
$\rho (x,z=L,t)=0$. The solution to eq. (\ref{Wave}) yields
\begin{equation}
\label{dr}
\frac{\dot{\rho}}{\bar{\rho}}\cong 2q\,\dot{\zeta} \, \frac{\sinh [p(z-L)]}{\sinh [pL]}
\end{equation}
with the familiar large-$p$ exponential decay
\begin{equation}
\label{drhighk}
\frac{\dot{\rho}}{\bar{\rho}}\Big{|}_{pL\gg 1}^{}\cong -2q\, e^{-pz}_{}\,\dot{\zeta}
\end{equation}
as well as the low-$p$ expansion
\begin{equation}
\label{drlowk}
\frac{\dot{\rho}}{\bar{\rho}}\Big{|}_{pL\ll 1}^{}\cong 2q\left(\frac{z}{L}-1\right)\dot{\zeta}\quad
\end{equation} 
determined by the cutoff length $L$ being smaller than the natural $1/p$ decay length. 
Alternatively, Eq. (\ref{drlowk}) can be obtained using the adiabatic approximation for slow low-$k$ modes, replacing
$f(z)$ in (\ref{rho}) with $f_{\mathrm{ad}}^{}(x,z,t)=(z-\zeta (x,t))/(L-\zeta (x,t))$.\\
In analogy, in the theory of the crystallization waves for He$_{}^{4}$ the induced velocity and density modulation in
the \emph{infinite} SF region decay as $\exp[-kz]$ and are proportional to the difference in the solid and liquid uniform
equilibrium densities. In our case the density difference is rather replaced by the "slope" $q$ since, if $q=0$ (i.e.
$\Delta\rho =0$) no fluid would move due to the displacement of the interface. 

Equation (\ref{contToy3}) thus has the general solution
\begin{eqnarray}
\label{psisol2} &&\psi (x,z,t)=\sum_{\sigma =\pm}^{} c_{\sigma}^{}(k)\,\exp\left[-Q^{}_{\sigma}(k)\, z\right]+ \\
&&+2q\dot{\zeta}\,\frac{(p^{2}_{}-k^{2}_{})\sinh [p(L-z)]+2qp\,\cosh [p(L-z)]}{\sinh [pL]
\left((p^{2}_{}-k^{2}_{})^{2}_{}-4p^{2}_{}q^{2}_{}\right)}\nonumber
\end{eqnarray}
with $Q^{}_{\sigma}(k)=q\left[1+\sigma \left(1+k^{2}_{}/q^{2}_{}\right)^{1/2}_{}\right]$. 
In (\ref{psisol2}) the coefficients $c_{\pm}^{}(k)$ are determined by imposing a vanishing
$v_{z}^{}=\partial_{z}^{}\psi$ at both $z=0$ and $z=L$. In the finite-$k$ limit this corresponds to oscillating
superfluid motion with non-vanishing $v_{x}^{}(z=0)$ and 
$v_{x}^{}(z=L)$. 
As a result, if we concentrate on slow surface waves ($p\simeq k$), the condition $qL\ll 1$ yields the expansions
\begin{eqnarray}
\label{vxvzexpand} 
v_{x}^{}\big{|}_{kL\ll 1}^{}&\cong&-i\,\dot{\zeta}\cdot\frac{q}{k}\; ,\quad 
v_{z}^{}\big{|}_{kL\ll 1}^{}\cong\dot{\zeta}\, q z\left(1-\frac{z}{L}\right) \nonumber\\
v_{x}^{}\big{|}_{kL\gg 1}^{}&\cong&-i\,\dot{\zeta}\cdot\frac{q}{k}\left(1+kz\right)\, e^{-kz}_{}\; ,\quad 
v_{z}^{}\big{|}_{kL\gg 1}^{}\cong\dot{\zeta}\, q z\, e^{-kz}_{}\nonumber
\end{eqnarray}
with the factor $i$ meaning a $\pi /2$ shift between the interface modulation $\zeta$ and the velocity $v_{x}^{}$. The
velocity is proportional to $\dot{\zeta}$ and becomes very large for small $k$. Thus, $\dot{\zeta}$ has to be thought of
as infinitesimally small to prevent the SF from exceeding the critical velocity after which superfluidity is destroyed.
\\
The kinetic energy per unit length of the interface can therefore be calculated as 
\begin{equation}
\label{Kin} K\cong\frac{1}{2}\, m\int_{0}^{L}\mathrm{d}z\, \bar{\rho}(z)\left|\mathbf{v}\right|^{2}_{}
\cong\frac{1}{2}\,\rho_{\mathrm{M}}^{}mL\frac{q^{2}_{}}{k^{2}_{}}\, G(kL)
\left|\,\dot{\zeta}\right|^{2}_{}
\end{equation}
with $G(x)$ a smooth function such that $G(x\ll 1)\simeq 1$ and $G(x\gg 1)\simeq 3/(2x)$. Its universal properties are
captured by the choice $G(x)=[1+(2x/3)^{2}_{}]^{-1/2}_{}$. 

From the potential energy (\ref{DeltaU}) and the kinetic term (\ref{Kin}) we derive the dispersion of the surface waves
\begin{equation}
\label{Disp}
\omega^{}_{} (k)\simeq \beta k\left[\left(1+\left(\frac{2}{3}\, kL\right)^{2}_{}\right)^{1/2}_{}\left(1 +
\left(\frac{k}{k^{}_{\Delta}}\right)^{2}_{}\right)\right]_{}^{1/2}
\end{equation}
with $\beta =\left[\Delta /\left(\rho_{\mathrm{M}}^{}mq_{}^{2}L\right)\right]_{}^{1/2}$ the group velocity in the low-$k$
linear regime and
$k^{}_{\Delta}=\left[\Delta /\left(\alpha (\hat{z})+\frac{
\partial_{}^{2}\alpha}{
\partial n_{x}^{2}}\right)\right]_{}^{1/2}$. For $k>1/L$ the interface at $z=L$ plays no major role and we recover the
$k^{3/2}_{}$ dispersion common to crystallization and capillary waves.

We now proceed to calculate the thermodynamic properties of the system, starting from the energy due to the 
occupation of the surface waves at temperature $T$,
\begin{equation}
\label{E} E=\hbar_{}^{2}\int_{0}^{\infty}\mathrm{d}\omega\,
\omega\rho_{}^{(\mathrm{1D})}(\hbar\omega)\frac{1}{e_{}^{\hbar\omega /K_{\mathrm{B}}^{}T}-1}
\end{equation}
with $K_{\mathrm{B}}^{}$ the Boltzmann constant and $\rho_{}^{(\mathrm{1D})}(\hbar\omega)$ 
the one-dimensional density of states. 
In the regime where the modes have a dispersion $\omega^{}_{}\sim
\gamma_{}^{}k_{}^{\alpha}$ this is 
\begin{equation}
\label{DOS}
\rho_{}^{(\mathrm{1D})}(\hbar\omega)=\frac{L_{x}^{}}{2\pi}\cdot\frac{1}{\alpha\left(\hbar\gamma\right)^{1/\alpha}_{}}\,
\left(\hbar\omega\right)^{1/\alpha-1}_{}
\end{equation}
with $L_{x}^{}$ the interface length ($L_{x}^{}\gg L$). The resulting contribution to the heat capacity
$C_{\mathrm{v}}^{(\mathrm{1D})}=
\partial_{T}^{}E$ is
\begin{equation}
\label{cv} C_{\mathrm{v}}^{(\mathrm{1D})}=\frac{L_{x}^{}}{2\pi}\cdot\frac{\Gamma (1+1/\alpha)\,
Z(1+1/\alpha)}{\alpha\left(\hbar\gamma\right)^{1/\alpha}_{}}\,
K_{\mathrm{B}}^{}\left(K_{\mathrm{B}}^{}T\right)^{1/\alpha}_{}
\end{equation}
with $\Gamma$ and $Z$ the Euler and Riemann functions. This has to be compared with the bulk contribution due to the SF
modes with $\omega^{}_{}= c\, (k^{2}_{x}+k^{2}_{z})^{1/2}_{}$. 
The wavenumber quantization along $z$ gives $k^{}_{z}=n\pi /L$ ($n\in \mathbf{N}$), yielding several 1D branches 
with energy split $E^{}_{g}=\hbar c \pi/L$ for vanishing $k^{}_{x}$. For
$K^{}_{\mathrm{B}}T\gg E^{}_{g}$ the bulk behaves as two-dimensional with 
\begin{equation}
\label{cv2D} C_{\mathrm{v}}^{(\mathrm{2D})}=\frac{L_{x}^{}L}{2\pi}\cdot\frac{6Z(3)}{\hbar ^{2}_{}c^{2}_{}}\,
K_{\mathrm{B}}^{3}T^{2}_{}\quad .
\end{equation}
For $K^{}_{\mathrm{B}}T< E^{}_{g}$, only the $n=0$ branch is involved in the heat capacity, yielding a contribution
analogous to (\ref{cv}) with $\alpha =1$ and $\gamma$ replaced by $c$.\\
Thus, in the low-temperature regime, the heat capacity is dominated by the 1D contribution and is linear in $T$. 

Before analyzing the implications of the results above, we estimate the typical parameters in our calculations, starting
with $L$. 
The phase diagram for the infinite system shows that the transition between a MI with on-site occupation $n$ and the
adjacent one with $n+1$, for a given ratio $J/U$, implies a \emph{maximum} jump $\mu /U =1$, independent on $n$. Thus, 
the effective chemical potential $\mu (z)=\mu -V(z)$ undergoes a jump of the order of $U$ between neighboring Mott-phases.
If
we describe the external confinement with the linear function $V(z)=\eta \, z$ this implies that the typical separation
between one MI and the next is $L\sim U/\eta$.\\
As far as $\Delta$ is concerned, we estimate the \emph{variation} of energy due to a small static density fluctuation
$\rho (\mathbf{r})$
\begin{equation}
\label{deltaE}
\delta U=\frac{1}{2}\int \mathrm{d}\mathbf{r}\,\mathrm{d}\mathbf{r}^{\prime}_{}\, \rho
(\mathbf{r})U(\mathbf{r}-\mathbf{r}^{\prime}_{})\rho (\mathbf{r}^{\prime}_{})
\end{equation}
with the on-site boson interaction $U(\mathbf{r}-\mathbf{r}^{\prime}_{})\sim U
l^{2}_{}\delta(\mathbf{r}-\mathbf{r}^{\prime}_{})$. In the $k=0$ case, via the adiabatic density fluctuation used in 
(\ref{drlowk}) and from eq. (\ref{DeltaU}) we get $\Delta  =U/(3 l^{2}_{}L)$.\\
Last, we are left with $\alpha (\hat{z})$ while the contribution $\partial_{}^{2}\alpha /
\partial n_{x}^{2}$ is neglected as for quantum rough surfaces. 
We consider a straight interface as an ideal line separating lattice sites in the MI phase from those in the SF one.
Neighboring sites in opposite phases are called "links", and the surface energy per unit length is the product of the
density of links ($1/l$) times the energy per link $J$ given by the hopping term in (\ref{H}), leading to $\alpha
(\hat{z}) =J/l$.\\
Out of our previous definitions and estimates we finally get $k^{}_{\Delta}=\left[\eta/3lJ\right]_{}^{1/2}$ as well as 
$\beta =\sqrt{2/3\pi}\, c$.\\
In the limit of large occupations $n_{\mathrm{M}}^{}\gg 1$ the critical ratio $(J/U)_{\mathrm{crit}}^{}$ needed to drive
the system into the SF-phase scales as $1/n_{\mathrm{M}}^{}$. Thus, in particular, we also need $U >
n_{\mathrm{M}}^{}J$ in order to have interfaces at all. In parallel, if the potential energy difference between
neighboring sites is equal to the MI gap $U$, resonant tunnelling destroys the insulating phase \cite{Greiner02}. 
Thus, in order to preserve the Mott phases, the confinement potential has to fulfill the condition $\eta\, l\le U$. 
These inequalities produce $k^{}_{\Delta}L\gg 1$. \\
%%%%%%%%%%%%%%%%%%%%%%%%%%%%%%%%%%%%%%%%%%%%%%%%
In the limit $J/U\ll 1$
and if the potential is steep, it is possible to achieve the regime $L\le l$. Here the SF rings disappear, yielding
neighboring MI domains whose stability was analyzed recently \cite{DeMarco}. In this case the ground state is
everywhere incompressible with energy gap $U$ and an activated heat capacity.
%%%%%%%%%%%%%%%%%%%%%%%%%%%%%%%%%%%%%%%%%%%%%%%%

As a result, while the bulk heat capacity is linear in $T$ up to $E^{}_{g}/K_{\mathrm{B}}^{}$ and grows as
$T^{2}_{}$ for high temperatures,
the contribution due to the 1D surface waves is linear in $T$ up to $\hbar \beta /L K_{\mathrm{B}}^{}$, it scales as
$T^{2/3}_{}$ between
$\hbar \beta /L K_{\mathrm{B}}^{}$ and $\hbar\omega (k^{}_{\Delta})/K_{\mathrm{B}}^{}$ to finally grow as $T^{2/5}_{}$ for $K_{\mathrm{B}}^{}T\gg
\hbar\omega (k^{}_{\Delta})$.\\
Thus, a direct measurement of the low-temperature heat capacity of the system will probe both the
\textit{existence} and the \textit{typical dimensions} of the SF rings. Indeed, the very fact that SF domains exist
introduces the cutoff length $L$ responsible for the transverse discretization of the bulk spectrum and the low-$k$
linear dispersion of the
surface waves. As a result, a linear-$T$ dependence of $C^{}_{\mathrm{v}}$ signals the existence of MI-SF domains.\\
Otherwise, if the atomic cloud is in a pure SF phase we expect a bulk $T^{2}_{}$ scaling down to very low temperatures.
Even in this case there is an extremely small quantization energy $E^{}_{0}$ associated to the lowest harmonic mode with
half wavelength equal to the total cloud diameter. For $K_{\mathrm{B}}^{}T < E_{0}^{}$ the system behaves as
zero-dimensional with an exponentially activated heat capacity.\\
%%%%%%%%%%%%%%%%%%%%%%%%%%%%%%%%%%%%%%%%%%%%%%%%%%%%%
Further informations about the number and typical size of the SF domains 
could be extracted from the \textit{slope} of the linear-$T$ part of $C_{\mathrm{v}}^{}$. Indeed, the total
heat capacity is given by summing the bulk and interface terms up to the number of SF domains $N$,
\begin{equation}
\label{cvlinear} 
C_{\mathrm{v}}^{(\mathrm{lin})}\simeq \,
\frac{\Gamma (2)\, Z(2)\, K_{\mathrm{B}}^{2}\, T}{2\pi\hbar}\,
\sum_{n=1}^{N} L_{x}^{}(n)
\left(\frac{1}{c_{n}^{}}+\frac{2}{\beta_{n}^{}}\right)\quad .
\end{equation}
Since the sound velocities $\beta_{n}^{}$ and $c_{n}^{}$ in the $n$-th domain can be controlled via the external
parameters, the slope of
$C_{\mathrm{v}}^{(\mathrm{lin})}$ yields informations about $N$ and the typical length of the SF rings.
Using typical experimental parameters indicated in \cite{Greiner02} we can estimate the temperature
$T_{g}^{}=E_{g}^{}/K_{\mathrm{B}}^{}$, below which we expect the 1D behaviour, to be of the order of 
$T_{g}^{}\sim 200\; nK$, offering a wide window for experimental detection of this effect.
%%%%%%%%%%%%%%%%%%%%%%%%%%%%%%%%%%%%%%%%%%%%%%%%%%%%%%%

In conclusion, after calculating the dispersion of the collective excitations localized at the MI-SF interfaces, we
deduced their contribution to the thermodynamic properties of a confined 2D Bose system on a lattice.
A direct measurement of a linear low-T heat capacity would signal the
\textit{existence} of MI-SF domains and its slope would tell about their \textit{typical size} via the external
control on $J$, $U$ and the confining potential. At higher temperatures, in presence of domains, a crossover to other
power-laws ($C_{\mathrm{v}}^{}\sim T_{}^{2/3},\, T_{}^{2/5}$) is to be expected, before the SF bulk $T_{}^{2}$ scaling
dominates the thermodynamic properties of the system. The latter is the only leading contribution to
$C_{\mathrm{v}}^{}$ in the absence of MI-SF domains, for a pure superfluid phase.\\

We acknowledge useful discussions with Ehud Altman and Yuval Oreg.
The support from the Feinberg School of the Weizmann Institute of Science, the US-Israel BSF and the Minerva
foundation is gratefully acknowledged.

\end{document}